\documentclass{iopart}

\usepackage{upgreek}
\usepackage{isomath}
\usepackage{siunitx}
\usepackage{graphicx}
\usepackage[latin1]{inputenc}

\begin{document}

\title{High repetition rate sub-ns electron pulses from resonant pulsed Rydberg field-ionization}

\author{R Hahn$^{1}$\footnote{Now at: Department of Chemistry and Applied Biosciences, ETH Zurich, CH-8093 Zurich, Switzerland} and D Comparat$^1$}

\address{Universit\'e Paris-Saclay, CNRS,  Laboratoire Aim\'e Cotton, 91405, Orsay, France.}
\ead{daniel.comparat@universite-paris-saclay.fr}
\date{\today}

\begin{abstract}
Using Resonant Pulsed Rydberg Field-ionization (RPRFI) technique, we generate low-energy electron bunches at high repetition rates. 
By combining continuous-wave laser excitation with a pulsed electric field, this method selectively ionizes Rydberg-Stark states in cesium atoms, producing  sub-ns long electron bunches (down to $\sim\SI{100}{ps}$) at a repetition rate of $\sim\SI{10}{MHz}$.
The method is demonstrated to offer significant advantages in terms of flexibility in the ionization repetition rate and pulse delay adjustments. The RPRFI method holds promise for applications in high-resolution electron microscopy and spectroscopy, potentially for overcoming the limitations of traditional electron sources in terms of brightness and energy spread.
\end{abstract}
\noindent{\it electron source, Rydberg states, pulsed field-ionization\/}

\section{Introduction}
Electron sources are essential to many applications in electron microscopy, electron spectroscopy, material sciences, nanotechnology, and biology \cite{woodruff2016modern,mozetivc2018recent,polman2019electron,king_ultrafast_2005,spence2019springer,rosenzweig2019next}.
Traditional electron sources such as thermionic and field-emission guns face limitations such as low brightness and high energy spread \cite{williams2009electron}. Consequently, improvements of the existing electron sources opens up new possibilities in their application \cite{musumeci2018advances}.

Some applications like high-resolution electron energy loss spectroscopy require low electron energy spread on the order of a few meV to resolve bulk/surface phonons or molecular vibrational modes. Attaining such low energy spread requires monochromatization of an electron beam from a conventional electron source like tungsten filaments, LaB$_6$ crystals, or field emission guns \cite{carsky2011low}, which produce electron beams with energy spread of $\sim \qty{250}{meV}$ at best.
Recent advancements in electron microscopy have also integrated monochromators in high-energy electron (transmission) microscopy columns \cite{krivanek2019progress}. Although this allows high-resolution spectroscopy and imaging in the same instrument, it entails high instrumental complexity and cost \cite{tromp2019spectroscopy,Lagos.2022}. Monochromators have additional drawbacks: they significantly attenuate the beam current and, for low energy applications, can increase angular divergence to such an extent that Low-Energy Electron Microscopy (LEEM) imaging conditions for surface characterization can no longer be fulfilled. 
As such, it would be advantageous to have a high-brightness electron source with low-energy width, as suggested in \cite{egerton2011electron}.
Such a source would allow the construction of a High-Resolution Electron Energy Loss Microscope \cite{mankos2019design,mcculloch2016cold} that would be able to perform LEEM while gathering spectroscopic information on the sample by Time-of-Flight (ToF) energy analysis.
The ToF method has the advantage of allowing the simultaneous detection of all electrons that will produce the image of the sample, without needing a slit to select a portion of the image (as would be the case for a hemispherical energy analyzer). 
This means microscopy and spectroscopic information can be gathered at the same time thanks to time- and position-sensitive detectors. 
Using ToF for energy analysis requires the electron source to be pulsed, with a temporal spread $\Delta \tau$ low enough to attain the required energy resolution. 

Electron sources based on the ionization of an atomic beam/cloud appear particularly promising for generating intrinsically monochromatic electron beams \cite{Gallagher.1974,mcculloch2016cold}. Their low emittance means that they can be bright enough to perform electron microscopy \cite{hahn2021ionization,mankos2019design}.
Most previous electron sources based on the ionization of an atomic beam were continuous \cite{Gallagher.1974,Klar.1994,Kurokawa.2010} but, using pulsed lasers, the generation of pulsed electron beams from these sources has also been reported \cite{rella1999high,zolotorev2010ultra,Engelen.2013,McCulloch.2013,fedchenko2020narrow,de2023subpicosecond}. However, pulsed lasers limit the selectivity of the excitation and do not allow for the selection of a particularly suitable state for electron emission.

In this work, we propose a novel method for generating low-energy electrons at high repetition rates, based on the ionization of individual Rydberg-Stark states of cesium atoms. 
This study builds upon our previous investigations into the field-ionization of Rydberg states for the production of ion or electron beams \cite{kime2013high,moufarej_forced_2017,mcculloch2017field,fedchenko2020narrow,hahn2021cesium}.
By combining a continuous-wave laser and a predominantly DC electric field with a minor oscillating component, we selectively excite and ionize Rydberg-Stark states of Cs atoms. 
This process results in the emission of $\sim \SI{100}{ps}$ to $\sim \SI{1}{ns}$ long electron bunches at a \SI{16}{MHz} repetition rate.  

In section \ref{sec:principle} we detail the working principle of the source, section \ref{sec:expe} presents the description of the experimental setup, and the results are given in section \ref{sec:results}, where we discuss the potential of this method for creating  monochromatic electron bunches with low temporal spread and low transverse-energy spread at a high-repetition rate.

\section{Ionization scheme: Resonant Pulsed Rydberg Field-ionization (RPRFI)}\label{sec:principle}
Upon exposure to a strong to moderate electric field $F$, different $l$-value Rydberg states of an atom mix and lose their characteristic $s$, $p$, or $d$ behavior in the field but remain resolvable.
Cesium, having the largest quantum defect among all stable alkali metals, displays exaggerate behaviors due to strong deviations from hydrogenic characteristics. 
Some of its states have substantial ionization rates even below the classical ionization threshold owing to the scattering of the Rydberg electron with the core. 
The Resonant Pulsed Rydberg Field-ionization (RPRFI) technique leverages our comprehensive investigation of this region of Rydberg excitation and ionization in Cs to identify suitable states for use in an electron source \cite{hahn2021cesium}.

\begin{figure}
    \centering
    \includegraphics[width=0.8\linewidth]{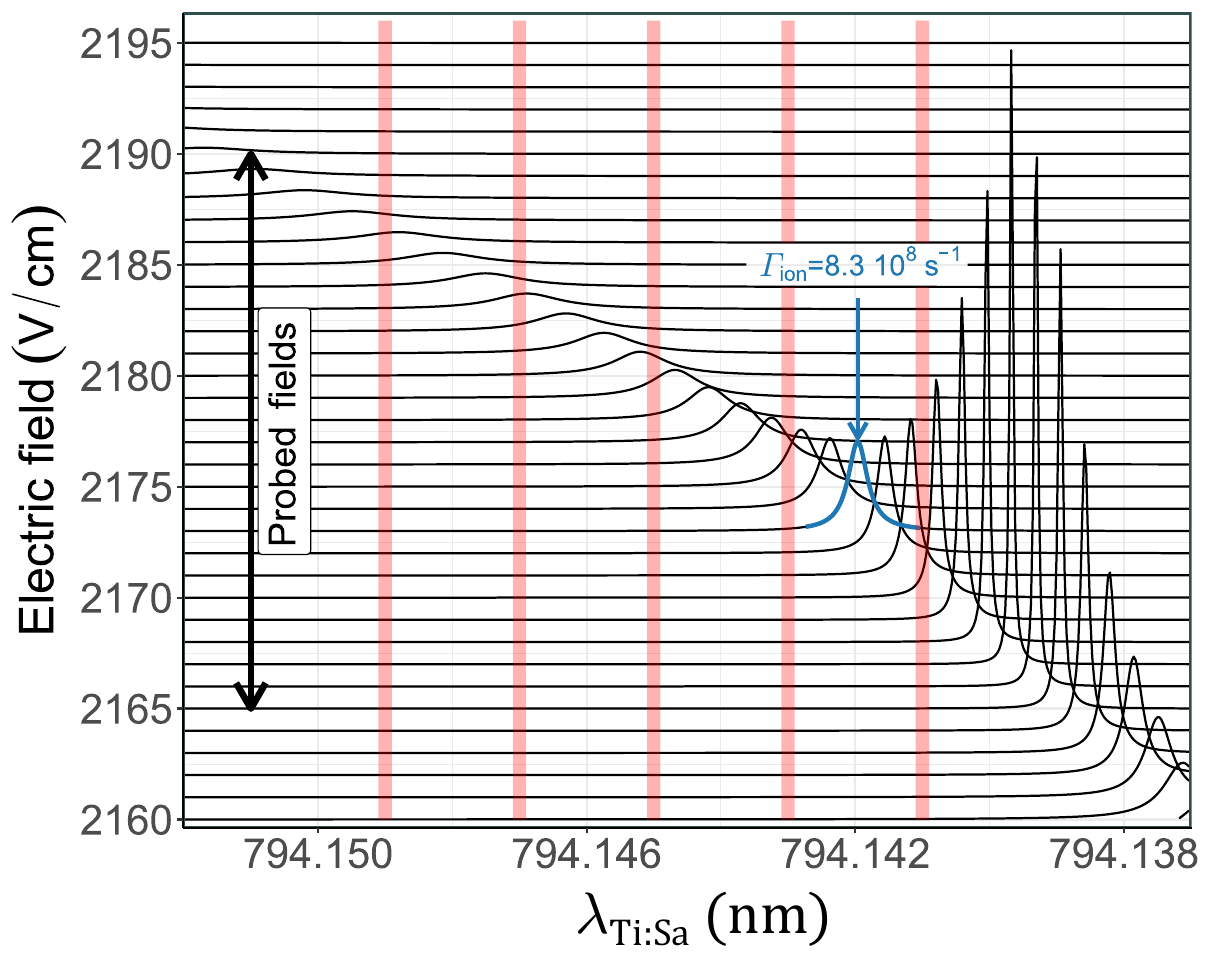}
    \caption{Photoabsorption spectra from the $7s\,(F=4)$ hyperfine level of cesium in electric fields between \SI{2160}{V/cm} and \SI{2195}{V/cm} calculated using the local-frame-transformation theory \cite{hahn2021cesium,Harmin.1982,Harmin.1982b}. 
    The $y$-offset indicates the electric field. The spectra show a transition to a Rydberg-Stark state with a strong Stark-shift (Stark-slope of $\frac{\rmd\nu}{\rmd F} \approx \SI{0.2}{GHz/(V/cm)}$) and field-dependent ionization rate.
    The red vertical lines represent the values of $\lambda_{\textrm{Ti:Sa}}$ at which the data in figure \ref{fig:result} was collected.
    The experimentally applied oscillating field of amplitude \SI{25}{V/cm} that brings the laser in and out of resonance is represented by a black double-arrow. 
    }
    \label{fig:StarkMap}
\end{figure}

For example, we identified regions of the Stark map where isolated Rydberg-Stark states, such as the one shown in figure \ref{fig:StarkMap}, possess a significant Stark effect and a strong variation of their linewidth with the electric field/laser wavelength, enabling us to investigate the relationship between the natural lifetime of the state and the observed ionization duration.
This photoabsorption Stark map was computed using local frame transformation (LFT) theory and WKB-wavefunctions as detailed in \cite{hahn2021cesium}.
This Rydberg-Stark state, given its strong Stark effect, can be excited by a fixed laser frequency $\nu$ only when the electric field $F$ is precisely tuned to resonance. We target Rydberg states with $n\approx 30$.
After excitation, the atom rapidly evolves into an electron and ion pair. To produce short electron bunches at a high rate, we pulse a small fraction ($\sim 10\%$) of the electric field, thereby bringing the laser in and out of resonance at a fixed and controlled frequency.
Each time the field brings the Rydberg-Stark level in resonance with the laser, an electron bunch is produced. The lifetime of this state is limited by its ionization rate $\Gamma_\text{ion}$, so the linewidth of the transition displayed on figure \ref{fig:StarkMap} gives us directly access to $\Gamma_\text{ion}(F)$.
Apart from $\Gamma_\text{ion}(F)$, the process's key parameters are the Stark slope of the Rydberg-Stark energy level $\frac{\rmd \nu}{\rmd F}$, and the electric field slew rate (SR) $\frac{\rmd F}{\rmd t}$. 
We will focus on these parameters in our analysis, discussing their relationship with the temporal spread of the electron bunches.
Assuming typical values for $\Gamma_\text{ion}$ close to \SI{1e9}{s^{-1}} (see figure \ref{fig:StarkMap}), $\frac{\rmd \nu}{\rmd F}$ around $\qty{100}{MHz/(V/cm)}$ and $\frac{\rmd F}{\rmd t}$ of a few $\qty{}{(V/cm)/ns}$, we can estimate the FWHM temporal width $\Delta \tau_{\rm FWHM}$ of the electron bunch,
\begin{equation}
\Delta \tau_{\rm FWHM} \approx  \frac{\Gamma_\text{ion}}{2 \uppi}   \left(  \frac{\rmd F}{\rmd t} \frac{\rmd \nu}{\rmd F}\right)^{-1},
\label{eq:timeSpread}
\end{equation}
to be in the ns range.

\section{Experimental methods}\label{sec:expe}
\subsection{General setup}
\begin{figure}
    \centering
    \includegraphics[width=\linewidth]{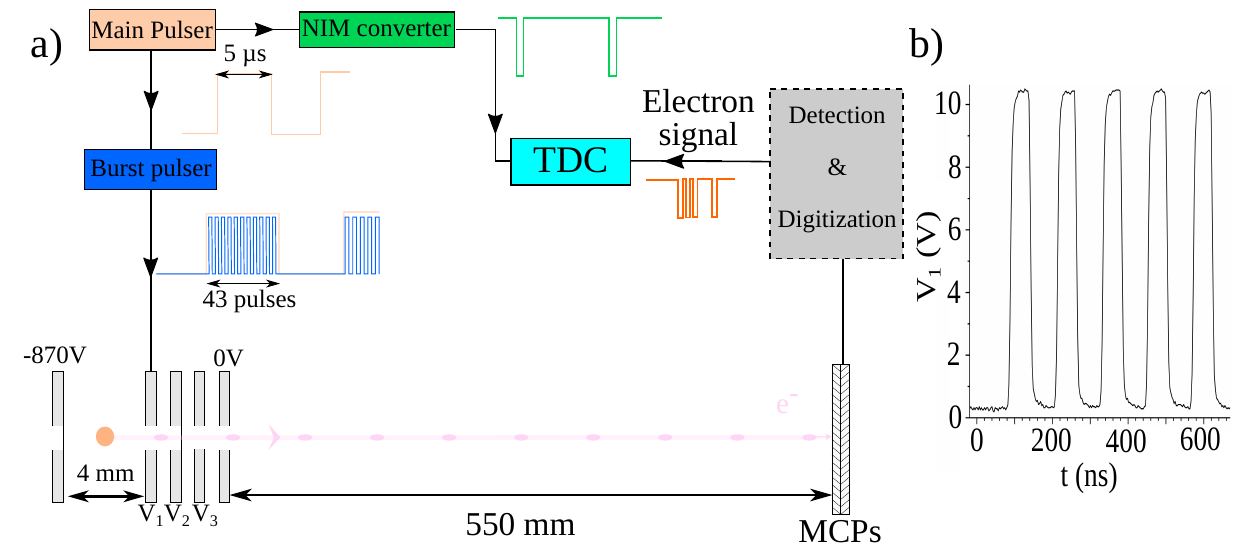}
    \caption{a) Diagram of the acquisition setup for pulsed resonance experiments. Electrode 0 (on the left) is maintained at -870 V and a potential $V_1$ (between \SI{0}{V} and \SI{10}{V}) is applied to electrode 1. This results in a field of approximately \SI{2175}{V/cm} between the two electrodes. Voltages $V_2$ and $V_3$ are adjusted to produce a converging electron beam exiting the electrode stack. The orange spot indicates the laser position.  
    b) Experimental recording of the time variation of the voltage $V_1$ applied to electrode 1.}
    \label{fig:setup}
\end{figure}

Figure \ref{fig:setup} presents the experimental setup, which closely mirrors that used in our prior studies \cite{moufarej_forced_2017} with the difference that now the Cs beam enters the setup laterally to mitigate Cs contamination on-axis. In brief, an effusive, recirculating oven (see \cite{hahn2022comparative} for details), produces a beam of atomic Cs that penetrates an electrode stack where two plates (electrodes 0 and 1) separated by approximately \SI{4}{mm} establish a DC electric field of around \SI{2200}{V/cm}.

Rydberg-Stark excitation is performed with three different $cw$-lasers 
 with wavelengths: $\sim\SI{852}{nm}$ for the transition $6p_{3/2}\,(F=5)\leftarrow 6s\, (F=4)$, $\sim\SI{1470}{nm}$ for $7s\,(F=4)\leftarrow 6p_{3/2}\,(F=5)$, and $\sim\SI{794}{nm}$ for $Rydberg\leftarrow 7s\,(F=4)$. Laser powers were respectively $\sim 10\,\mu$W, $\sim 100\,\mu$W, and \SI{10}{mW}, chosen to minimize power-broadening of the lines.
The first two laser radiations (\SI{852}{nm} and \SI{1470}{nm}) are produced by DL Pro and DFB Butterfly laser diodes from Toptica. The Rydberg-excitation laser is a Ti:Sa laser, and its wavelength is monitored using a HighFinesse WS8-2 high-precision wavemeter. During data acquisition, all three lasers are frequency-locked; we use a standard vapor cell-based method for the first two lasers, and the wavemeter for the Ti:Sa laser. The waists of the lasers are on the order of tens of micrometers, and the last two lasers (\SI{1470}{nm} and Ti:Sa) cross orthogonally.

An oscillating electric field (see next section) drives the laser in and out of resonance of the excitation towards a Rydberg-Stark state. 
Rydberg-Stark states with short ionization lifetimes are excited and produce electrons. 
These electrons are accelerated by the electric field and focused by two additional electrodes (denoted by $V_2$ and $V_3$ on figure \ref{fig:setup}) onto a double stack microchannel plate (MCP), followed by a phosphor screen placed approximately \SI{55}{cm} downstream.
Electrode 0 is kept at \SI{-870}{V}, producing a potential of around \SI{-435}{V} where the electrons are created. These electrons are accelerated to \SI{435}{eV} at the end of electrode 4, and so at the front of the MCPs, which are kept at \SI{0}{V} (see figure \ref{fig:setup} a)).

\subsection{Electron time-of-flight measurement}

As our primary interest lies in the electron pulse duration, we implemented a detection method with approximately \SI{60}{ps} resolution, based on a high-performance Time-to-Digital Converter (TDC).
The main clock of the experiment is a commercial pulser that provides long pulses ($5\,\mu$s) and triggers the opening of the TDC. 
These pulses trigger another pulser that produces a burst of shorter pulses with individual width of \SI{50}{ns}, rising times of \SI{10}{ns}, and an amplitude of $V_1=\SI{10}{V}$. These shorter pulses are directed to electrode 1 via a short, non-shielded cable, to minimize the pulse time-broadening during propagation.

The oscillating electric potential on electrode 1, shown in figure \ref{fig:setup} b), creates an oscillating electric field in the laser excitation region.
With electrodes 0 and 1 separated by \qty{0.4}{cm}, a \SI{10}{V} pulse on electrode 1 induces an electric field pulse of amplitude \SI{25}{V/cm}.
Electrons are detected by standard detectors (MCPs), and the signal they produce is amplified and digitized by a Constant Fraction Discriminator (CFD).
This signal is subsequently transmitted directly to a time-to-digital converter (TDC: TDC-V4 from the DTPI platform), which concurrently records trigger pulses from the main pulser. 
A schematic of the acquisition setup is presented in figure \ref{fig:setup}. 
The TDC opens independently for 50\,$\mu$s, registering two types of ``stop" signals during this period (with a binning time of \SI{125}{ps}): pulses from the \textit{main} pulser and electron hits from the MCPs. Data is accumulated for a few minutes and recorded by a dedicated acquisition software.
The software saves the delay between each electron hit and the preceding nearest main pulse signal, enabling the reconstruction of time-of-flight spectra from a multitude of unique and uncorrelated events. 

\section{Results}\label{sec:results}
\begin{figure*}
    \centering
    \includegraphics[width=\linewidth]{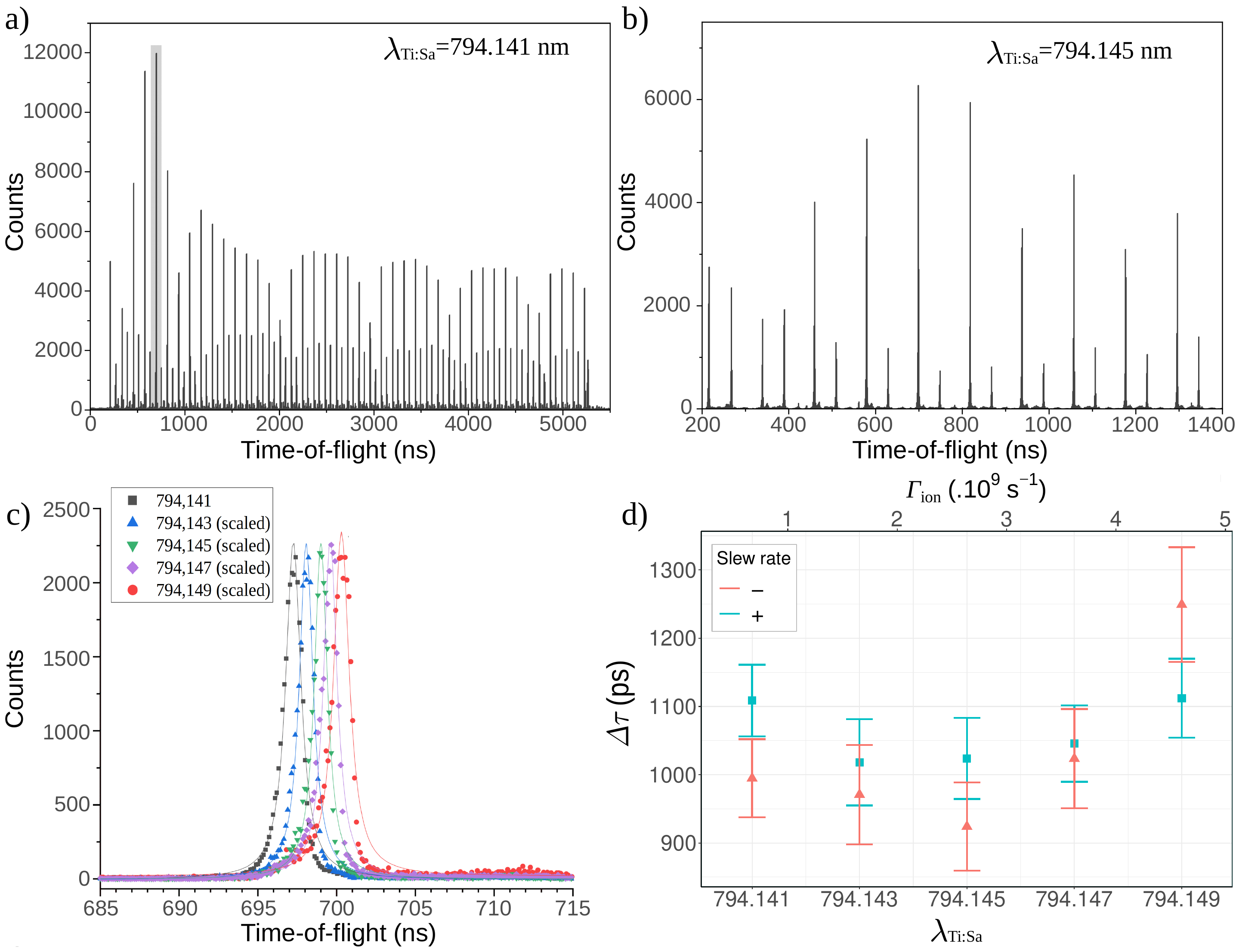}
    \caption{a) Complete electron ToF spectra with $\lambda_{\textrm{Ti:Sa}}=\SI{794.141}{nm}$.
    b) Zoomed-in view on a few peaks of the ToF spectrum acquired at $\lambda_{\textrm{Ti:Sa}}=\SI{794.145}{nm}$.  
    c) Shape and positions of the most intense peak in scans taken at five different laser wavelengths. The data at \SI{794.141}{nm} is taken as an intensity reference and all other data are scaled to it.
    d) Electron temporal spread (Lorentzian FWHM) of the 32 first pairs of peaks in the spectrum as a function of $\lambda_{\textrm{Ti:Sa}}$ (and thus $\Gamma_\text{ion}$, cf. figure \ref{fig:StarkMap}) and of the sign of the SR (see text for details). Positive and negative SRs are indicated in blue and red, respectively. Error bars represent the standard deviation.
    }
    \label{fig:result}
\end{figure*}

Each time the electric field brings the laser in resonance, electrons are generated, both on the rising and falling fronts of the electric pulse applied on electrode 1, leading to doublet structures in the electron signal.
A typical ToF spectrum for $\lambda_{\textrm{Ti:Sa}}=\SI{794.141}{nm}$ is shown in figure \ref{fig:result} a), displaying series of electron bunches. 

A few doublets from the data with $\lambda_{\textrm{Ti:Sa}}=\SI{794.145}{nm}$ are shown in detail on figure \ref{fig:result} b). The first peak of each doublet (corresponding to the rising front) is more intense than the second, with the separation between these two peaks corresponding to the voltage pulse length, which can thus be freely tuned by tuning the repetition rate of the burst pulser.

\subsection{Influence of $\Gamma_\text{ion}$ on $\tau_{\textrm{ion}}$}\label{sec:Gamma}
We studied experimentally the influence of the state's linewidth on the observed bunch duration. Data was collected on the state depicted in figure \ref{fig:StarkMap} under the same conditions except for the tuning of the Ti:Sa's wavelength, which was adjusted from \SI{794.141}{nm} to \SI{794.149}{nm} in increments of \SI{0.002}{nm}.
The Stark map of figure \ref{fig:StarkMap} establishes a link between the laser wavelength, the electric field, and the state linewidth. 
In particular we take advantage of the quasi-linearity between the studied state's ionization rate $\Gamma_\text{ion}$ and its resonant laser wavelength to study the influence of $\Gamma_\text{ion}$ on  $\Delta \tau$.

Figure \ref{fig:result} c) shows the most intense peak of the electron ToF spectrum for each value of $\lambda_{\textrm{Ti:Sa}}$ (gray rectangle on panel a)), on the same time axis, fitted with Lorentzian peak functions.
We fit similarly the first 32 pairs of peaks in the spectrum with Lorentzian peak functions and report the FWHM as the temporal width $\Delta\tau$ on figure \ref{fig:result} d), separating peaks with positive SR (electrons created during the rise of the electric field pulse) from the peaks with negative SR (electrons created during the fall of the electric field pulse). The mean values are around \SI{1}{ns}, with a minimum mean value around \SI{930}{ps} for $\lambda_{\textrm{Ti:Sa}}=\SI{794.145}{nm}$ and a negative SR.
The relation between the state's ionization rate $\Gamma_\text{ion}$ and the electron bunch temporal length seems to suggest an optimal ionization rate $\Gamma_\text{ion}=\SI{2.7e9}{s^{-1}}$, but, surprisingly, the influence of $\Gamma_\text{ion}$ on the ionization dynamics appears to be quite limited.

According to equation \ref{eq:timeSpread}, the SR should have a large influence of the observed electron bunch duration. Changing the SR in a controlled way is experimentally quite challenging, but we noticed that our experimental SR was actually not constant during the measurement, allowing us to study the influence of the SR on $\Delta\tau$.

\subsection{Influence of the slew rate on $\tau_{\textrm{ion}}$}\label{sec:slewrate}
\begin{figure}
    \centering
    \includegraphics[width=0.6\linewidth]{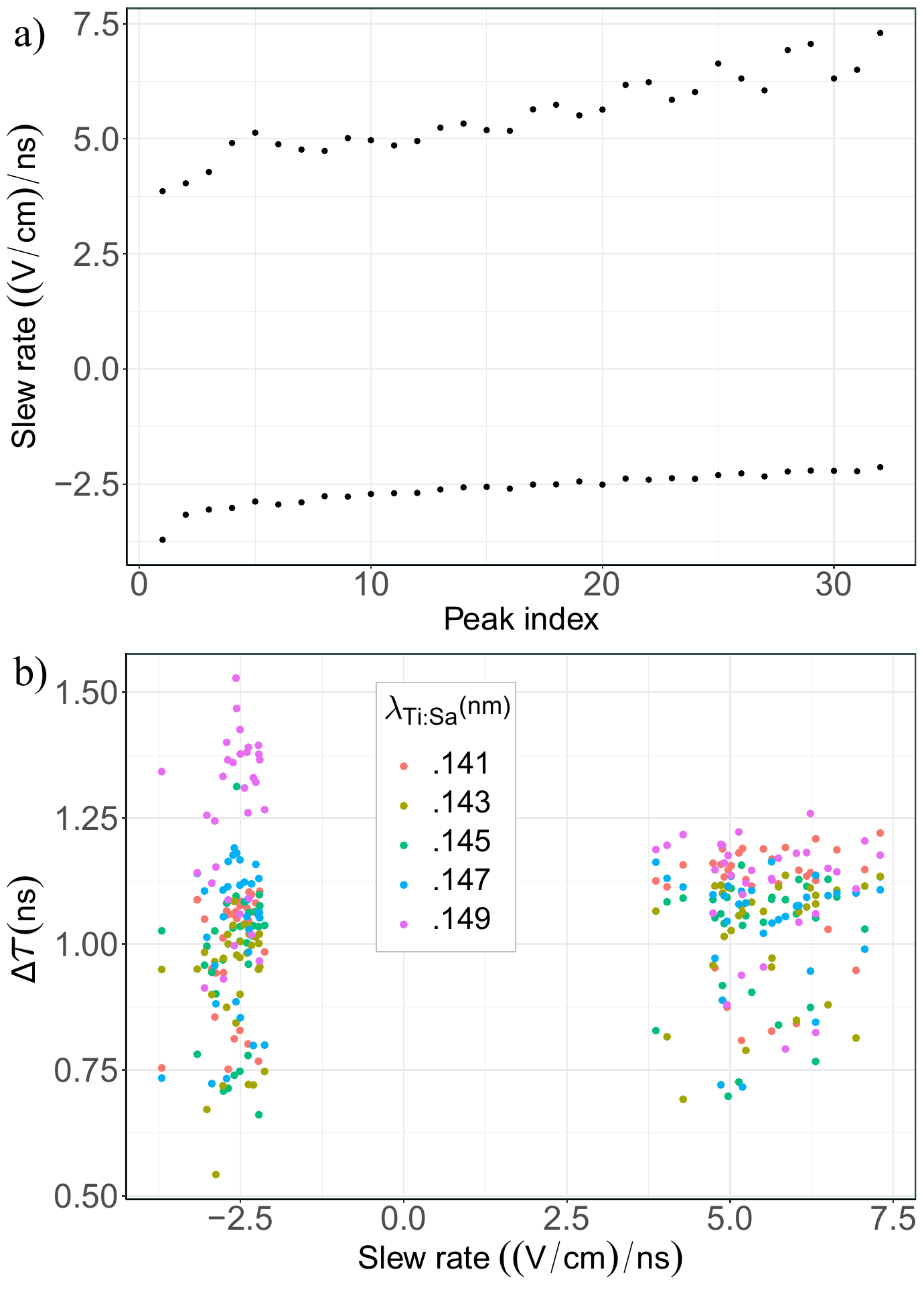}
    \caption{a) Slew rate extracted from the data presented on figure \ref{fig:result} for the first 32 pair of peaks (see text for details). The positive SR are about twice as large and increase during the series of pulses, while the negative SRs decrease (in amplitude) but at a smaller rate.
    b) Lorentzian FWHM of the electron bunches emitted in function of the SR of all fitted peaks, for all five values of $\lambda_{\textrm{Ti:Sa}}$ (indicated in color), showing an absence of clear correlation.
    }
    \label{fig:slewrate}
\end{figure}

As visible on figure \ref{fig:result} c), the central ToF of the peaks decreases linearly with the laser wavelength. Because the Stark shift is also linear (see figure \ref{fig:StarkMap}) and that other experimental conditions are kept constant, we attribute this time delay to the slightly different resonant electric fields that are reached at slightly different times, i.e. to the SR.
Together with the Stark slope given by the Stark map $\frac{\text{d}\nu}{\text{d}F}$,
the change in the central ToF $\frac{\text{d}\nu}{\text{d}t}$ provides a value for the experimental SR $\frac{\text{d}F}{\text{d}t}$:
\begin{equation}
    \frac{\text{d}F}{\text{d}t}=\left(\frac{\text{d}\nu}{\text{d}F}\right)^{-1} \frac{\text{d}\nu}{\text{d}t}.
\end{equation}
From the fitted position of each peak for the five different values of $\lambda_{\textrm{Ti:Sa}}$ we thus determine the SR for each peak, and plot them on figure \ref{fig:slewrate} a). We see clearly that the positive SRs are about twice as large as the negatives ones. The positive SRs increase significantly in amplitude over time, while the negative SRs decrease slightly in amplitude. 
The evolution of the measured $\Delta \tau$ in function of the SR is visible on figure \ref{fig:slewrate} b) (data with different values of $\lambda_{\textrm{Ti:Sa}}$ are indicated in different colors). Again, the influence of $\frac{\text{d}F}{\text{d}t}$ on $\Delta \tau$ seems to be very small, with no clear correlation between these two values.
The FWHM temporal spread seems to be limited to about \SI{1}{ns} for this state.

\subsection{Data at $\lambda_{\textrm{Ti:Sa}}=\SI{793.933}{nm}$}\label{sec:i0}
Another set of data was gathered, in a different region of the Stark map, with the intent to generate shorter electron bunches. The idea is to excite a state that has a relatively small ionization rate, just before an avoided crossing that increases rapidly its ionization rate.
Such a state near \SI{2065}{V/cm} can be seen on figure \ref{fig:rh1} a), its position in the field highlighted in blue.
The laser frequency was set to \SI{793.933}{nm} (see red vertical ribbon) and the field was oscillating between \SI{2055}{V/cm} and \SI{2070}{V/cm} so that the excitation happens during the field pulse. The resulting electron pulses are shown on figure \ref{fig:rh1} b), which shows a large (\SI{3.4}{ns}) and weak pulse followed by a very intense and narrow peak (Lorentzian FWHM of \SI{287\pm20}{ps}) for the rising electric field. The falling edge of the electric field pulse shows no peak (expected around \SI{770}{ns}).
The first peak is attributed to the state visible around \SI{2060}{V/cm}, and the second, very sharp peak is attributed to the state indicated in blue on the Stark map.
The absence of bunches on the falling edge and the one-order-of-magnitude difference in the $\Delta \tau$ between the two states are difficult to explain from the Stark map alone and are yet to be understood.
These features demonstrates once more that the electron dynamics leading to ionization can be very complex, especially in parts of the spectrum where several states are close in energy. The ionization dynamics can not simply be described by the state's ionization rate, and a more in-depth, theoretical analysis is required here, for example as done in reference \cite{rella1999high}. This complexity allows for example to generate asymmetric but regular electron pulses from symmetric electric field pulses. 
Furthermore, it is remarkable to see that an electron pulse with such a narrow temporal width can be created with this method.
We now investigate whether the measured temporal width of $\SI{287}{ps}$ is representative of the limit of the method, or of our measurement.

\subsection{The effect of energy spread on the measured temporal spread}
The measured temporal spread $\Delta \tau$ is indeed a convolution of $\tau_{\textrm{ion}}$ and $\tau_{\Delta E}$, where $\tau_{\textrm{ion}}$ is the intrinsic duration of the ionization process, and $\tau_{\Delta E}$ is the temporal spread caused by the energy spread.
For energy-spread dominated bunches, $\Delta \tau$ is strongly dependent on the position of measurement: for instance sub-ps bunches have been reported where the measurement is performed at the so-called self-compression point, which occurs near twice the acceleration distance \cite{de2023subpicosecond}.
However self-compression cannot yield a temporal spread smaller than the emission time $\tau_{\textrm{ion}}$, as it merely cancels out the spatial (and hence, temporal) elongation of a bunch locally.

\begin{figure}
    \centering
    \includegraphics[height=\textheight]{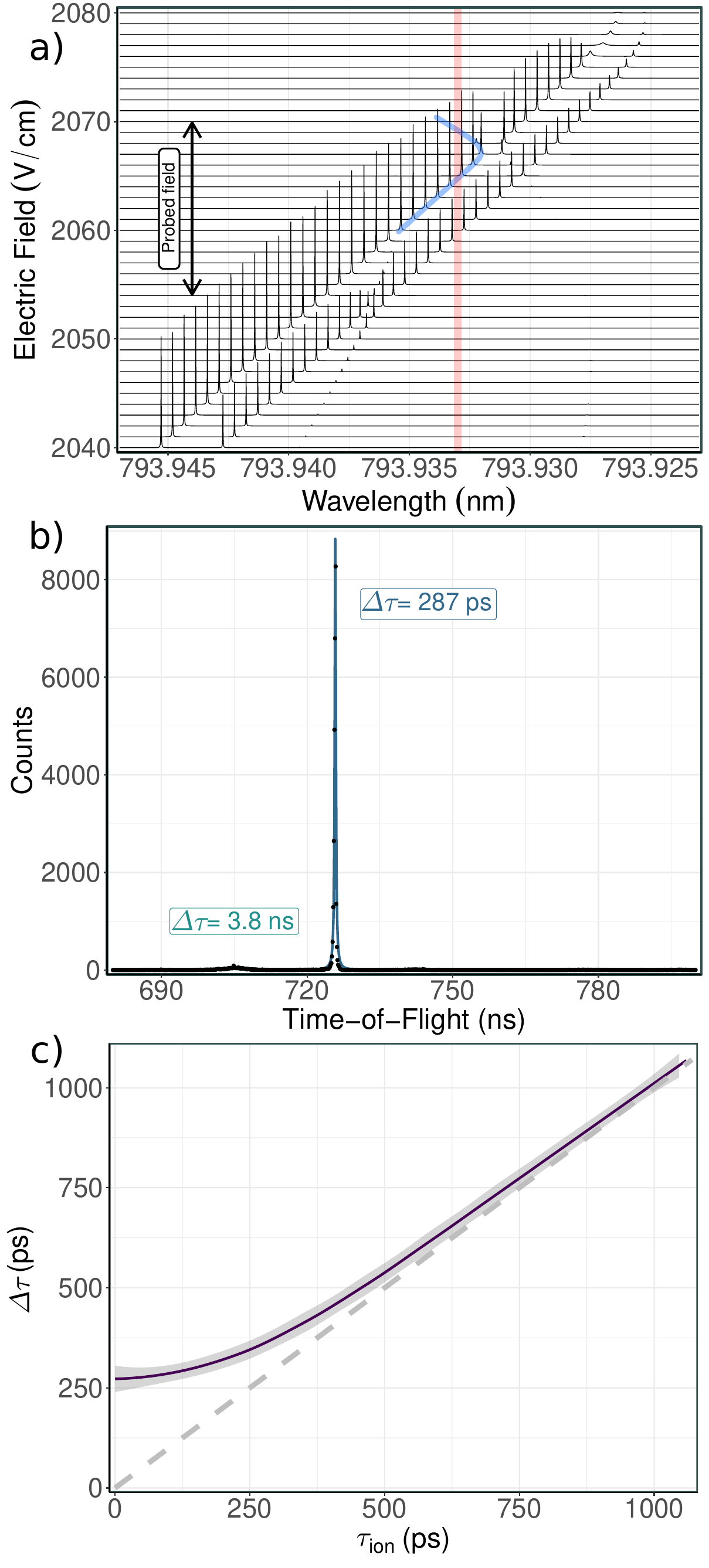}
    \caption{a) Stark map with $\lambda_{\textrm{Ti:Sa}}=\SI{793.933}{nm}$ indicated in red and $F$ pulsed from \SI{2055}{V/cm} to \SI{2070}{V/cm} indicated by a double-arrow.
    b) Resulting electron pulses with Lorentzian FWHM of \SI{3.8}{ns} and \SI{287}{ps} corresponding to the rising edge of the electric field. 
    c) Evolution of the measured time spread $\Delta \tau$ in function of the ionization time $\tau_{\textrm{ion}}$, taking into account the energy spread of the electron source and our experimental conditions. The width of the line represent the scatter of the simulations results. The $y=x$ line is shown as a dashed gray line to guide the eye.
    }
    \label{fig:rh1}
\end{figure}

To assess the influence of the electron bunch energy spread on the measured temporal spread, we conducted simulations using the \textsc{SIMION} software \cite{Dahl.2000}.
We compared the arrival temporal spread $\Delta \tau$ for electron bunches with identical $\Delta E$, but with varying $\tau_{\textrm{ion}}$. $\Delta E$ is chosen to best describe our experimental conditions, i.e. a Gaussian distribution of FWHM close to $\SI{14}{eV}$.
The results are shown in figure \ref{fig:rh1} c), showing clearly that the energy spread limits the measured temporal spread to be above \SI{280}{ps}.
Because the temporal width presented in figure \ref{fig:result} is around $\SI{1}{ns}$, we can be confident that the measured $\Delta \tau$ reflects the intrinsic $\tau_{\textrm{ion}}$.
On the contrary, for the data presented in figure \ref{fig:rh1} b), the measured $\Delta \tau$ is below \SI{300}{ps}, so that $\tau_{\Delta E}$ has a large influence on $\Delta \tau$. The main consequence is that $\tau_{\textrm{ion}}$ is very probably around or even below \SI{150}{ps}. 

Reducing the energy spread (by focusing the laser more tightly or using lower fields) would thus also significantly reduce the measured time spread. 
Calculations suggests that similar states are available at much lower field, for example around \SI{10}{V/cm}, where a laser focused to \SI{10}{\micro m} would create electron bunches with $\Delta E\approx\SI{10}{meV}$. Experiments are under way to find these states and verify that the temporal width are similar to the ones measured in this work.
In terms of transversal energy spread, previous measurements of ion/electron correlations have shown \cite{hahn2021ionization} that the transverse energy spread of these Rydberg-based electron source are on the order of \SI{1}{meV}, which is comparable or even better than the best electron sources from cold atoms \cite{Engelen.2013,franssen_compact_2019}.

\section{Conclusion}
We have demonstrated the use of the RPRFI technique to create electron bunches at a high repetition rate ($\sim \SI{16}{MHz}$) with a pulse duration on the order or even well below the \si{ns} range. The lowest width we have measured is at \SI{287\pm20}{ps}, limited by our measurement procedure.
Using a cw-laser allows Rydberg-Stark state-selective excitation and ionization, even in the strongly mixed region of the Stark map, thereby allowing fine tuning of the emission properties of the source by targeting appropriate states.
Shaped electron bunches can be produced with this methods, provided that an appropriate part of the Stark map is used.

We studied experimentally the influence of the states' ionization rate and of the experimental slew rate on the temporal spread of the electron bunch and found almost negligible influence.
One notable advantage of replacing a pulsed laser with a pulsed electric field is the ability to freely adjust the ionization repetition rate and the delay between pulses by modifying the delays in voltage pulses. 
This is much simpler than adjusting repetition rates or delays in an optical system and can very easily produce electrons at a higher rate if required (up to $\sim \SI{100}{MHz}$).

The electric field that we employed (around \SI{2170}{V/cm}) currently limits the energy spread (FWHM) to approximately \SI{14}{eV}. This limitation is due to the voltage difference among electrons generated in different parts of the cloud, which is a result of the product of the electric field by the cloud size.
Thus, in order to further decrease the energy spread of our source and reach $\Delta E<\SI{10}{meV}$ that would open vibrational spectroscopic possibilities \cite{mankos2019design}, we will need to decrease the electric field by two to three orders of magnitude, and limit the size of the ionization zone by focusing the laser more tightly.

The lowest temporal width that we report is already at the time-resolution of next-generation detectors for electron microscopes, like the TimePix4 \cite{Ballabriga2023}. For applications that would require smaller temporal width, active bunching could be considered, particularly techniques that allow electron bunch compression without increasing their energy spread \cite{hahn2021ionization}.
By simply reversing the fields, this source could also provide ions in a focused ion beam or ion-implantation studies \cite{kime2013high}.
Additionally, the correlation between the produced ions and electrons can be leveraged in different ways to improve the properties of the source \cite{Lopez2019,hahn2021ionization}. 
Finally spin-polarized electrons could be easily produced by these sources, especially when using cesium \cite{Fano.1969,Mollenkamp.1982}.

Therefore, our high repetition-rate pulsed low-energy electron source based on atomic Rydberg ionization could find applications in various fields, including materials science, surface science, and biological imaging.

\ack
This work was supported by the Fond Unique Interminist\'eriel (IAPP-FUI-22) COLDFIB, the labex PALM (grant number ANR-10-LABX-0039-PALM) and the ANR-HREELM project (ANR-14-CE35-0019).

 A CC-BY public copyright license has been applied by the authors to the present document and will be applied to all subsequent versions up to the Author Accepted Manuscript arising from this submission, in accordance with the grant's open access conditions.

We acknowledge N. Barrett and L. Amiaud for fruitful discussions and Y.J. Picard for sharing knowledge and access to the data acquisition setup.

\end{document}